\documentstyle[12pt]{article}
\begin{document}



\vskip 3em

\Large
\centerline{\bf  Atomic form factor }
\centerline{\bf  for twisted, vortex photons}
\centerline{\bf interacting with atoms}

\normalsize

\vskip 2em

\centerline{\it Pierson Guthrey, Lev Kaplan and J. H. McGuire} 
\centerline{\it Department of Physics and Engineering Physics} 
\centerline{\it Tulane University, New Orleans, LA 70118}

\vskip 3em
{\bf Abstract:}  The relatively new atomic form factor for twisted (vortex) beams,  which carry orbital angular momentum (OAM), is considered and compared to the conventional atomic form factor for plane wave beams that carry only spin angular momentum (SAM). 
Since the vortex symmetry of a twisted photon is more complex that that of a plane-wave,  evaluation of the atomic form factor is also more complex for twisted photons.  On the other hand, the twisted photon has additional parameters, including the OAM quantum number, $\ell$, the nodal radial number, $p$, and the Rayleigh range, $z_R$ that determines the cone angle of the vortex.  This Rayleigh range may be used as a variable parameter to control, in new ways, the interaction of twisted photons with matter.  
Here we address: i) normalization of the vortex atomic form factor, ii) displacement of target atoms away from the center of the beam vortex, and iii)  formulation of transition probabilities for a variety of photon-atom processes.  We attend to features related to new experiments that can test the range of validity and accuracy of calculations of these variations of the atomic form factor.  Using the absolute square of the form factor for vortex beams, we introduce a vortex factor that can be directly measured.  


\vskip 2em
email:  mcguire@tulane.edu or lkpalan@tulane.edu

\newpage

\section{Introduction}
The atomic form factor is used to calculate observable count rates for processes involving the interaction of macroscopic beams interacting with submicroscopic targets, such as atoms. This atomic form factor, $M$, enables scattering into directions that differ from scattering when transitions in atomic target states are absent.  This factor itself describes the relative strengths of the various transitions that couple beams, e.g. of photons, electrons or ions, under various scattering conditions.   While it is now conventional to describe these transitions in terms of propagating plane-waves, one may also use other basis states, such as, for example, states that describe propagation of twisted photons\cite{ Allen92,YaoPadgett}.   
Although calculations of atomic form factors for plane wave beams\cite{TU, Chantler00, nist} have been available for some time\cite{nist, GarveyGreen76}, relatively few calculations\cite{AfanasevEtal, MatulaEtal,DKM} are now available for twisted (vortex) beams, even though such twisted beams themselves have been available since 1936  \cite{Beth}.  

In recent years there has been an increasing number of experiments exploring the nature of twisted photons\cite{YaoPadgett,Molina-TerrizaTorresTorner, BabikerBADR, MolinaTerrizaRTTW, BerryMcDonald,GiovanniniNMS}, as well as twisted (vortex) electrons\cite{IvanovSerbo, Ivanov}, that have followed mathematical descriptions of vortex beams in both a Laguerre Gaussian basis, applicable in the paraxial approximation\cite{Allen92, YaoPadgett},  as well as a less restrictive Bessel basis\cite{MatulaEtal, IvanovSerbo}.  There are widely varied  applications\cite{Molina-TerrizaTorresTorner,TorresTornerBook} that use twisted vortex beams, including quantum information\cite{WildeUskov, BarreiroWeiKwiat, MairVWZ}, quantum control\cite{Bucks, Shakovetal},  optical tweezers\cite{PadgettBowman}, Bell inequalities\cite{MTetal, AltmanEtal} and vortex phenomena in Bose-Einstein condensates\cite{AndersenEtal}, as well as broader applications in other fields\cite{Molina-TerrizaTorresTorner}.

Twisted (vortex) beams carry orbital angular momentum\cite{YaoPadgett}.  Orbital angular momentum (OAM) differs from spin angular momentum (SAM).  Plane wave photons carry only the more widely studied SAM, while twisted photons carry both SAM and OAM, which arises from the vortex nature of the twisted photon beam.
Plane wave photons may be characterized in terms of two variable mathematical parameters, $(\lambda, m_s)$, where $\lambda$ is the (continuously variable) photon wavelength and $m_s$ is the quantum number describing (two) possible azimuthal directions of the intrinsic photon spin, $s = 1$.
Twisted photons may be characterized\cite{DKM} in terms of five variable mathematical parameters, $(z_R, \lambda, p, \ell, m_s)$, where $z_R$ is the (continuously variable) Rayleigh range of the twisted photon that sets the angle of the vortex cone, $p$ gives the number of nodes in the photonic radial distribution, and  $\ell = \pm |\ell|$ describes the magnitude and direction of the orbital angular momentum, $\hbar |\ell|$.  
Twisted photons have a more restricted symmetry than plane-wave photons.  However, the constraint of this adjustable twisted vortex can act as a macroscopic control handle for submicroscopic atoms.  Comparison of the possible values of $(\lambda, m_s)$ with $(z_R, \lambda, p, \ell, m_s)$ suggests that more control over atomic processes may be achieved with twisted, vortex photons than with plane wave photons\cite{Molina-TerrizaTorresTorner}. 

In this paper we detail the nature and utility of the atomic form factor for twisted photons, and relate the twisted form factor to new experiments on interactions of vortex beams with matter.  
Here we consider beams described in terms of a traveling plane wave, $e^{i (\vec{k} \cdot \vec{r} - \omega t)}$ modified by a Laguerre-Gauss profile\cite{Allen92,YaoPadgett}.  Our attention is centered here on photons in the optical regime interacting with hydrogenic atoms.  The photons each carry momentum $\vec{p} = \hbar \vec{k}$ and energy $E = \hbar \omega$.

This paper follows an earlier paper\cite{DKM} that formulated the interaction of twisted photons with atoms and gave an equation for the atomic form factor for twisted photon beams.  Here we address three significant unresolved issues in this earlier paper.  First, the expression for the atomic form factor in the earlier paper was not properly normalized, although the difficulty in doing so was traced to a difference in the normalization for plane waves and for twisted photons.   Understanding the different ways one may normalize amplitudes for systems with different symmetries (as well as different size scales) enables comparison of theory with experiment.   Second, it was assumed that the target atom was at the center of the beam vortex.  The effect of moving the atom away from this vortex was left for future study.  This question has also been recently considered by Afanasev et al.\cite{AfanasevEtal} and Matula et al.\cite{MatulaEtal}.   We show how understanding the effect of moving a target atom away from center of the incoming beam enables design of experiments using macroscopic gas cell.  Third, the authors called for a formulation of the scattering problem in position space to calculate  probability amplitudes using the atomic form factor.  This was also left for future study.  Here we detail the transition probability amplitude, expressed in position space, which is complementary to the scattering amplitude, expressed in momentum space, and which is more widely used in analysis of optical processes.  
We also address some additional considerations related to experiments exploring the processes involving interactions of twisted beams with atoms, including a way to isolate and observe the effect of vortex size.

\section{Atomic form factors}

\subsection{SAM form factors}

The atomic form factor describing the interaction of plane-wave beams with an atomic target\cite{DKM,WikiAF} is,

\begin{eqnarray}
     M = <f| e^{i \vec{q} \cdot \vec{r}} |i> = \int \phi^*_f(\vec{r})  e^{i \vec{q} \cdot \vec{r}} \phi_i(\vec{r}) d^3\vec{r} \ \ ,
  \label{M}
\end{eqnarray}
 \noindent
where $\phi_i$ and $\phi_f$ are the wavefunctions for the initial and final states of the atomic target, and $\hbar \vec{q} = \hbar \vec{k}_i - \hbar \vec{k}_f$ is the momentum transfer of the scattered photon. This plane-wave atomic form factor, $M$, describes the quantum transition from one atomic state to another with a possible change in photonic SAM, but bypasses activity in OAM.   This SAM form factor appears in a variety of physical processes, including photo-absorption, photo-emission, and photon scattering\cite{Mc}. 
Tables of $|M|^2$ are used for applications in which specific atomic transitions are significant, including applications involving photons interacting with matter\cite{TU, nist, Chantler00,GarveyGreen76}. 

For photons interacting weakly with atoms, where first order perturbation methods are applicable, one may usually simply multiply a classical differential cross section by $|M|^2$ to obtain the corresponding process when the target undergoes a transition from an initial state, $|i>$ to a final state, $|f>$.  For example, the differential cross section for Compton scattering is given by\cite{Heitler, Mc1}, 
\begin{eqnarray}
  d \sigma_C/d\Omega = |M|^2 \ d \sigma_T/d\Omega  \ \ , 
   \label{dsig}
\end{eqnarray}
where,
$ d \sigma_T/d\Omega = \frac{\omega_f}{\omega_i} r_0^2 |\hat{\Lambda_i} \cdot \hat{\Lambda_f}|^2 $
is Thomson's classical differential cross section for scattering of light from a free electron. 
Here $\hat{\Lambda_i}$ and $\hat{\Lambda_f}$ are the directions of the polarization of initial and final photons and $r_0 = 2.818 \times 10^{-15}$m, is the classical radius of the electron, and the relativistic Klein-Nishina relativistic correction term\cite{KleinNishina} has been omitted.  
As an aside we mention that any additional interaction between beam and the target, not included in the square of the canonical momentum of the electron-photon system is, we conventionally assume, small.
The quantum mechanical scattering cross section differs from its corresponding classical cross section when $|M|^2 \neq 1$.

\subsection{OAM atomic form factors}

The atomic form factor, $M_v$, for vortex photons describes the transition from one atomic quantum state to another using a mathematical basis describing incoming and outgoing twisted photons. 
While some calculations involving the atomic form factor for twisted photons have been recently published\cite{AfanasevEtal, MatulaEtal,DKM}, to our knowledge none has been in a form that may be compared directly to experiment, in part due to unresolved issues described in our introduction.  

An unnormalized expression for the atomic form factor, $M_v$, for the interaction of vortex photons with atoms has been given in an earlier paper\cite{DKM}, where the $ e^{i \vec{q} \cdot \vec{r}}$ factor for plane-waves is modified by a Laguerre-Gauss profile that describes the vector potential for twisted photons within the paraxial approximation\cite{Allen92}.  After multiplying Eq.(7) of the earlier paper by a factor of $\frac{1}{2} \lambda z_R$ so that $M_v$ reduces properly to the plane-wave limit,  the expression for the dimensionless, scale invariant atomic form factor becomes, 

\begin{eqnarray}
M_v  = \frac{1}{2} \lambda z_R  \int d\vec{r} \,  \phi_{N_f,L_f,M_f}^*(\vec{r}) \ u_{p_f,  \ell_f}^*(\vec{r}) \ e^{-i \vec{k}_f \cdot \vec{r} } 
 \ e^{i \vec{k}_i \cdot \vec{r} }  
 u_{p_i, \ell_i}(\vec{r}) \phi_{N_i,L_i,M_i}(\vec{r})    
  \label{MtGeneral}
\end{eqnarray}
\begin{eqnarray}
=   \frac{1}{2} \lambda z_R  \int d\vec{r}  \,   \phi_{N_f,L_f,M_f}^* (\vec{r})       
 \sqrt{\frac{2p_f!}{\pi (p_f + |\ell_f |)!}} \frac{1}{w(r \cos\theta')} 
 \left[\frac{ r|\sin \theta'| \sqrt{2}}{w(r \cos\theta')}\right]^{|\ell_f | } 
\exp\left[ \frac{-(r \sin\theta')^2}{w^{2}(r \cos\theta')}\right]
\nonumber
\end{eqnarray}
\begin{eqnarray}
\times L_{p_f}^{|\ell_f |}\left(\frac{2 (r \sin\theta')^2}{w^{2}(r \cos\theta')}\right) 
e^{-i \ell_f \phi'}
\exp\left[ \frac{-i k_f (r \sin\theta' )^2 r\cos\theta'}{2((r \cos\theta') ^2 + z^{2}_R)}\right] 
\exp\left[i(2p_f + |\ell_f | +1) \tan^{-1}\left(\frac{ r \cos\theta' }{z_R}\right)\right]    
 \nonumber
 \end{eqnarray} 
 \begin{eqnarray} 
\times  e^{i\vec{q} \cdot \vec{r} }
 \sqrt{\frac{2p_i!}{\pi (p_i + |\ell_i |)!}} \frac{1}{w(r \cos\theta)} 
    \left[\frac{ r|\sin \theta| \sqrt{2}}{w(r \cos\theta)}\right]^{|\ell_i |} 
\exp\left[ \frac{-(r \sin\theta)^2}{w^{2}(r \cos\theta)}\right]
L_{p_i}^{|\ell_i |}\left(\frac{2 (r \sin\theta)^2}{w^{2}(r \cos\theta)}\right) 
\nonumber
\end{eqnarray}
\begin{eqnarray}
\times  e^{i \ell_i \phi}
\exp\left[ \frac{i k_i (r \sin\theta )^2 r\cos\theta}{2((r \cos\theta) ^2 + z^{2}_R)}\right] 
\exp\left[-i(2p_i + |\ell_i | +1) \tan^{-1}\left(\frac{ r \cos\theta }{z_R}\right)\right]  
    \phi_{N_i,L_i,M_i} (\vec{r}) 
\label{Mt}
\end{eqnarray}
In Eq.(\ref{MtGeneral}) $u_{p_f,  \ell_f}$ and $u_{p_i, \ell_i}$ are the Laguerre-Gauss factors that change a plane-wave photon, $e^{i \vec{k} \cdot \vec{r}}$, into a twisted photon, as specified in Eq.(\ref{Mt}), corresponding (in cylindrical coordinates, where $\rho = r \cos \theta$, $z = r \sin \theta$ and $\phi$ is unchanged) to\cite{Allen92,YaoPadgett}, 
\begin{eqnarray}
  u_{p, \ell}(\rho,z,\phi) = 
  \sqrt{\frac{2p!}{\pi (p+|\ell|)!}} \, \frac{1}{w(z)} \left [ \frac{\rho\sqrt{2}}{w(z)}\right ]^{|\ell|}\exp\left [ -\frac{\rho^2}{w^2(z)} \right ]L_{p}^{|\ell|}\left ( \frac{2\rho^2}{w^2(z)} \right ) \nonumber \\
 \times \exp[i\ell\phi]\,\exp\left [ \frac{i k \rho^2z}{2(z^2+z_R^2)} \right ]
 \exp\left [ -i(2p +|\ell| +1)\tan^{-1} \left ( \frac{z}{z_R} \right )\right ]   \  . 
\label{LG}
\end{eqnarray}
Here, the twisted photon carries momentum $\hbar \vec{k}$ propagating in the $\hat{z} = \hat{k}$ direction, $w(z) = w(0)\sqrt{1+z^2/z^2_R}$, and $w(0) = \sqrt{\lambda z_R/\pi}$ is a measure of the minimum width of the beam (beam waist) determined by the Rayleigh range, $z_R$ .  
The phase, $(2p + |\ell| +1) \tan^{-1}({z}/{z_R})$, is the Gouy phase, and $ L_p^{|\ell|}(x)$ is an associated Laguerre  polynomial related to the more familiar Laguerre polynomials by 
$ L_p^{|\ell|}(x) = (-1)^{|\ell|} \frac{{d}^{|\ell|}}{dx^{|\ell|}} L_{p+|\ell|}(x)  $. 
The index $p$ is the number of radial nodes between $p+1$ peaks in the intensity distribution, and $\ell$ is the azimuthal index.
The beam envelope at a fixed wave intensity is described by $w(z)$, depicted in Fig. 1.
In Eq(\ref{LG}), ($\theta$, $\phi$) are spherical angles defined so that the north pole $\theta=0$ corresponds to the $\hat{k}_i$ direction, while
$(\theta'$, $\phi')$ is a coordinate system rotated by the scattering angle $\Theta$, so that $\theta'=0$ corresponds to the $\hat{k}_f$ direction. 

Next we resolve some mathematical and conceptual issues raised earlier\cite{DKM}, as mentioned in our introduction.   While the example we use is transfer of OAM in elastic Compton scattering, most of our discussion applies to other processes, including those using electron (and ion) beams and some other targets, as well as enabling experimental efforts to assess the range of validity of the formula used to evaluate various atomic form factors for twisted photons. 

\subsection{Normalization of plane-wave and twisted-wave photon amplitudes}

The factor of $\frac{1}{2} \lambda z_R$ used above to normalize the formula for $M_v$ makes the atomic form factor dimensionless (by removing a factor of $\frac{1}{w(0)}$ used earlier\cite{DKM}), and for $\ell_i = \ell_f = p_i = p_f = 0$ it also normalizes the corresponding SAM atomic form factor to $\delta_{fi}$ as $q \rightarrow 0$  in the plane-wave limit of large $z_R$ when formulated in a Gaussian basis. 
The dimensionless atomic form factor can join vastly different size scales.  On the macroscopic (beam) scale the length scale is $w(0) = \sqrt{\lambda z_R/\pi}$, which is about $10^{-4}$ m for an optical photon with $\lambda \sim 5 \times 10^{-7} $ m and $z_R \sim 10^{-1} $ m.
On the submicroscopic (target) scale the length scale is the size of the atomic target, $a_T = N^2 a_0$, where $a_0 = 5.29 \times 10^{-11}$ m, is the Bohr radius and $N$ is the principal quantum number.  

The plane wave terms are volume normalized, while the twisted photon terms are area normalized, as we now briefly explain.
The plane wave vector potential is volume normalized, i.e., $\vec{A} =  \hat{\Lambda} \sqrt{ \pi e^2/ r_0 E V}  \ e^{i (\vec{k}\cdot \vec{r} - \omega t)}$, where $\hat{\Lambda}$ is the direction of polarization, $e$ is the electron charge, $r_0 = \frac{e^2}{m_e c^2}$ is the classical radius of the electron ($m_e$ is the electron mass), $E$ is the photon energy, and $V$ is the volume of the photon beam.  The vortex photon terms are normalized to the size of the beam radius, $w(z) = w(0) \sqrt{1 + z^2/z_R^2}$, at each value of $z$.
Within the paraxial approximation, which requires the ray trajectory to be approximately parallel to the beam axis (discussed in Sec. 2.5 and 2.6 below), the difference in the volume and area normalizations can be accounted for by a factor of $\int dz = cT$, corresponding to the length of the beam, needed to convert an area to a volume.  Since this length is usually quite large compared to the time interval of the atomic interaction, the $\int dz$ term conventionally gives rise to an energy conserving $\delta$-function in the derivation for the formula for the  wave-like scattering amplitude for this process\cite{Econs}.  In this case the interaction region may considered  to be point-like in both space and time on the macroscopic scale of the beam.    Thus the $\int dz$ term is conventionally absorbed into the scattering amplitude in a wave-like picture.  

\subsection{Effect of moving atoms away from an optical beam center}

Since optical photons have a wavelength, $\lambda$, much larger than the size of the atomic target, $a_T$, it is sensible to formulate scattering of optical photons from atoms in momentum representation and employ the scattering amplitude, $f(\vec{q})$.   In this wave picture, the momentum of the photon is well defined, as is the scattering angle, $\Theta$, between the incoming photon of  momentum, $\hbar \vec{k}_i$, and  the outgoing photon of momentum,  $\hbar \vec{k}_f$.  For conceptual and mathematical  simplicity we now take  $k_i = k_f = k$ so that  $q = | \vec{k}_i - \vec{k}_f | = 2 k \sin(\Theta/2)$.  Extension to inelastic collisions, where $k_i \neq k_f$, is straightforward\cite{Mc}, and is briefly mentioned again below in the section on experimental considerations.  

We emphasize, as discussed in increasing detail in the next subsections, that the momentum space scattering amplitude, $f(\vec{q})$,  is the Fourier transform of the position space probability amplitude, $a(\vec{b})$, that is commonly used when the beam projectile is regarded as particle-like, as is the case for many ion-atom collisions \cite{MC1}. 
Since the probability amplitude in position space is the Fourier transform of the scattering amplitude in momentum, the effect\cite{ArfkinWeber} of the factor $e^{i \vec{q} \cdot \vec{b} }$ in wave space is to shift the position space distribution by a distance $\vec{b}$.   Since the size of the atom is small compared to the photon beam parameters ($\lambda$ and $z_R$), it is sensible to regard the atom as point-like, so that $\vec{b}$ represents a translation from the center of the photon beam here taken to define the location of $\vec{b} = 0$.  That is, $\vec{b}$ describes a classical light ray that interacts with an atom located a distance $\vec{b}$ from the center of the photon beam.  We call $\vec{b}$ the impact parameter for for this photon-atom collision, where $\vec{b}$ is defined on the scale of the macroscopic beam.  

On the other hand if the photon were considered as point-like (e.g. if $\lambda$ and $z_R$ were both quite small compared to the target size, $a_T$), then it is sensible to regard the target as large, and the projectile can be described a point-like particle.  This is often the case for ion-atom collisions, for example.  Such a description can be sensible for high energy photons.  In this case of scattering of a particle from an atomic target, it is common to take $\vec{b}$ as the distance of a straight line particle trajectory from the center of a relatively large target.  This is the conventional definition of the impact parameter $\vec{b}$ for classical particle collisions\cite{Mc1a,Golds}.  In the particle-like picture of an incoming projectile, it is conceptually sensible to consider $\vec{b}$ as the displacement of the point-like projectile from the center of the target, now taken to define $\vec{b} = 0$.  Thus, the concept of the impact parameter $\vec{b}$, of a collision is different for classical particles and classical waves.  As we consider more fully below, when describing beams incoming to a target in the classical particle-like limit it is conceptually convenient to use the probability, $|a(\vec{b})|^2$, to describe a collision event, while in the classical wave-like limit it is convenient to use the square of the scattering amplitude, $|f(\vec{q})|^2$.  Quantum mechanics incorporates both of these particle-like and wave-like limits.  

As discussed even further in the next subsection, since $a(\vec{b})$ and $f(\vec{q})$ are position and momentum amplitudes related by Fourier transforms, Parseval's theorem requires mathematically that the total cross section cross section summed over $\vec{b}$ is the same as that summed over $\vec{q}$.  That is, the total number of events counted experimentally by summing over $\vec{q}$ (defining the solid angle in ($\Theta, \phi$)) is the same as that by summing over $\vec{b}$ (defined by $(b,\phi)$), where the azimuthal angle $\phi$ is common to both coordinate systems.   The impact parameter, $\vec{b}$, can mathematically be used similarly in both classical limits, but the conceptual meanings are physically different.

\subsection{Relation between quantum amplitudes in position and momentum representations}

As introduced above, there are two complementary ways to evaluate cross sections.  In the wave-like approach, widely used for beams of optical photons interacting with atoms, the differential cross section is expressed in terms of the scattering amplitude, namely, 
$ d^2\sigma/d\Omega = |f(\vec{q})|^2$, where $q = q(\Theta)$, e.g. $q = 2 k \sin(\Theta/2)$ for elastic scattering.   Alternatively, in the particle-like approach, used for ion-atom collisions\cite{Mc}, the differential cross section is expressed in terms of the probability amplitude, namely, $ d^2 \sigma/d^2b = |a(\vec{b})|^2$.  In quantum mechanics, the amplitude, $f(\vec{q})$ in momentum space is related to the probability amplitude, $a(\vec{b})$, by the 2D relation, $ a(\vec{b}) = \frac{1}{2 \pi k} \int e^{i \vec{q} \cdot \vec{b}}  f(\vec{q}) d\vec{q} $.  Since the momentum representation and the position representation are Fourier transforms of one another, it follows that\cite{Mc2},

\begin{eqnarray}
\int |a(\vec{b})|^2 b db \ d\phi =   \frac{1}{(2 \pi k)^2}  \int  |f(\vec{q})|^2 q dq \ d\phi = \sigma_{total}
\label{totsig}
\end{eqnarray}
To our knowledge, this was first discussed in the context of atomic collisions in 1966 by McCarroll and Salin\cite{MS}.

Since $f(\vec{q})$ is the Fourier transform of $a(\vec{b})$, one has the constraint that $\Delta b \Delta q \geq \frac{1}{2}$.  If $\Delta b $ is small compared to the target size, $a_T$,  then the projectile may be regarded as a particle traveling on a classical trajectory $\vec{R} = \vec{b} + \hat{z} v t$.  If $\Delta  q$ is so small that $\Delta b$ is large compared to the target size, then the projectile may be regarded as a classical wave directed along a ray perpendicular to the wavefronts.  
Since classical plane waves are invariant under translation perpendicular to the beam, and $e^{i \vec{q} \cdot \vec{r} }$ produces a translation in position space, then for plane-wave photons $a(\vec{b})$ must be independent of $b$, so that $a(\vec{b}) = a(0)$.  Indeed the dependence of $a(\vec{b})$ on $\vec{b}$ is simply disregarded in widely used books and papers dealing with photons.  In any case, the translation symmetry of plane-waves is broken by vortex waves.  
For any twisted beam (see Fig. 1) each ray trajectory, $\vec{R} = \vec{b} + \vec{v} t$, is a straight line, where $\vec{b}$ is the displacement of this trajectory from the center of the beam at $z = 0$, and $\hat{v}$ has both a $\hat{z}$ component and a component in the $x-y$ plane, as dicussed below.  (One may choose $\hat{x} = \hat{q}$, defining an $x - z$  plane of scattering.)  For light traveling in free space $v = c $, while for beams of electrons, protons and ions, $v < c$.   When scattering occurs this trajectory passes through the target atom in a semi-classical picture where ray trajectories are used.  It is the rotation of the trajectory  about the beam axis that provides orbital angular momentum to the beam.  This rotation is provided by $e^{i \ell  \phi}$ in the Laguerre-Gaussian terms in Eq(\ref{LG}) that give a parity breaking handedness to the twisted vortex beam. 

In our experience there are advantages to each method of calculation, i.e., using either $q$-space or $b$-space.  In many cases for both photon beams and particle beams, the momentum representation most accurately matches experimental conditions since it is the momentum of the scattered particle that is accurately observed.  Also in our experience the mathematics is a little simpler in the momentum representation.  On the other hand, using the position representation yields a dimensionless probability, $|a(\vec{b})|^2$, that may not exceed unity.  This upper limit can be useful in testing approximate theoretical calculations.  Also the probability directly indicates how likely a transition is under varying conditions.  This could be useful in applications involving the transfer of quantum information.  Moreover, this formulation, employing both $f(\vec{q})$ and $a(\vec{b})$, opens the way for new theoretical and experimental studies of dynamic correlation  in a wide range of systems, including scattering of photons from multi-electron atomic systems\cite{Mc}. 

\subsection{Relations involving $\vec{b}$}

In this subsection, we relate the parameter, $\vec{b}$, that appears in the position representation described above, to various physical quantities.  This subsection is intended to clarify the conceptual meaning of $\vec{b}$ under differing physical conditions.   

For a vortex Gaussian beam\cite{vortexClass}, the asymptotic beam intensity decreases exponentially as the distance, $\rho$, from the axis of the beam increases. In Eq.(\ref{LG}) when $\rho$ is equal to $w(z)$, for $\ell = p = 0$ the beam intensity drops by a factor of $\frac{1}{e^2}$ from its maximum occurring at $\rho = 0$.  As illustrated in Fig 2,  when $\rho(z) = w(z)$, the so called\cite{vortexClass} 'divergence angle of the beam', $\Theta_{\cal V}$, asymptotically approaches a constant value, since  $w(z)$ increases linearly with $|z|$  at large $|z|$ (see Fig 2C), namely,
  
\begin{eqnarray}
      \tan \Theta_{\cal V} = w(z)/|z|  \rightarrow w(0)/z_R  = \frac{\lambda}{\pi w(0)} \ \ .
   \label{b-ThetaV}
\end{eqnarray}
where $w(z) = w(0) \sqrt{ 1 + z^2/z_R^2} \rightarrow w(0) \frac{|z|}{z_R} $ when $|z| > > z_R$. 

Now $\rho(z)$ itself may vary from 0 to $\infty$.  Defining $\rho(z) \equiv \frac{b}{w(0)} \ w(z)$, for varying asymptotic angles, $\Theta_V(b)$ [that may differ from the particular value, $\Theta_{\cal V} = \Theta_V(b = w(0))$], one has,

\begin{eqnarray}
      \tan \Theta_V(b) = \rho/|z| = \frac{b  \ w(z)}{w(0) \  |z|}  \rightarrow b / z_R  \ \ .
   \label{b-Thetav}
\end{eqnarray}
Fig 2 shows $w(z)/w(0)$ depicting cylindrically symmetric intensity layering in a simple Gaussian beam ($\ell = p = 0$).  
Since the beam intensity at a thin ring of radius $\rho(z)$ about the $z$ axis is, at any $z$, the same as the beam intensity at the corresponding radius $\rho(0) = b$, the beam intensity encountered by a (small) atomic target displaced a distance $b$ from the center of the beam at $z = 0$ is the intensity at the asymptotic angle $\Theta_V(b)$.  
Thus, the asymptotic values of $\tan \Theta_V(b)$ of Eq(\ref{b-Thetav}), illustrated in Fig 2C, increase linearly with the distance from the center of a vortex beam, $b$.  At any single  value of $b$, every small atom confined  to the region $z \leq z_R$, will encounter a beam of approximately the same intensity since $w(z) \simeq w(0)$, as illustrated in Fig 2B.   
Thus, $w(0)$ is a physical length that determines the beam intensity in a thin ring a distance a distance $\rho(z) =  b \frac{w(z)}{w(0)}$ perpendicular to the axis of the vortex beam.

For Laguerre Gaussian beams with $\ell \neq 0$, the overall Gaussian envelope for the intensity is the same, but additional nodes associated with the quantum number, $p$, may appear from the Laguerre factors in Eq(\ref{Mt}).  For such a twisted vortex beam that carries OAM, the $e^{i \ell \phi}$ factor in Eq(\ref{Mt}) allows rotation about the beam axis of a point of constant phase.  Thus a classical light ray, which passes through any target atom which is displaced a distance, $b$, away from the beam axis at $z = 0$, may rotate and does carry OAM. This defines the (possibly rotating) classical light ray that passes through an atom displaced by $\vec{b}$ (perpendicular to $\hat{z}$) from the vortex center at $z = 0$.  
A classical light ray that follows a single, non-rotating trajectory is depicted by one of the straight line segments shown in Fig 1.
Near $z = 0$, the magnitude of the maximum OAM is classically limited by $\hbar \ell \leq \hbar b k$.  

In quantum collisions of twisted vortex beams with atoms many outcomes are possible, including transfer of OAM. Then the direction of photonic OAM is simply reversed. Transfer of OAM may occur only when $\ell \neq 0$ in the photon beam. 
For example, in transfer of OAM, where $\ell \rightarrow -\ell$ for the twisted photon, a corresponding change in the atomic orbital angular momentum is required to conserve parity.  In general a non-zero scattering angle, $\Theta$, provides linear momentum transfer to the atom via the momentum transfer, $\vec{q} = \vec{k}_i - \vec{k}_f$. The strength of the particular final state of the system is modulated by the $e^{i \vec{q} \cdot \vec{r}}$ factor in Eq(\ref{Mt}), as discussed previously\cite{DKM}.

We note that for all $\ell$ there is an angle, $\tilde{\chi}$, between $\vec{k}_i$ and the beam axis.  This may be most easily seen in Fig 1, where it is evident that  $\tilde{\chi} = \Theta_{\cal V}$ for an incoming beam.  For $b \neq w(0)$,  $\tilde{\chi} = \Theta_V(b)$ (see Fig 2).  In the paraxial approximation the effect of this rotation is small.  

Collisions of atoms with electron, proton and ion beams are different than collisions with photon beams. Here the classical particle limit often applies.  Then the position of the incoming projectile is well defined, and $\vec{b}$ corresponds to the impact parameter of the collision, i.e., the distance of closest approach of the particle to the center of a target for a classical, undeflected  straight line trajectory.  For a collision with Coulomb deflection between two charged particles the asymptotic beam angle, $\Theta_C$, is classically related to the magnitude of the impact parameter, $b$, by\cite{Golds},

\begin{eqnarray}
    \tan( \Theta_C/2) = d_0/(2b)   \ \ ,
   \label{b-ThetaC}
     \end{eqnarray} 
where $d_0$ is the distance of closest approach between the point-like projectile and the point-like target in a head on collision. 
In quantum calculations the parameter, $b$, in Eq(\ref{b-ThetaC}) is mathematically the same as that in Eq(\ref{b-Thetav}).
The angles $\Theta_C$ and $\Theta_V(b)$ have different meanings. The angle $\Theta_C$ is the difference between the incoming and outgoing asymptotic directions of one charged particle classically scattering from another charged particle.  It is a classical scattering angle.  The angle $\Theta_V(b)$ is related to the beam intensity, as explained below Eq(\ref{b-Thetav}) above.  
The Coulombic scattering angle, $\Theta_C$ is largely controlled at small distances where the Coulomb force is strong, while the vortex angle, $\Theta_V(b)$, is controlled at large distances from the center of the vortex.  
We note that in many high velocity electron-atom collisions, most projectiles are scattering into forward angles and the incoming electron may be treated as a plane wave to a good approximation and Coulomb deflection may be ignored.  

In a full quantum description of a process with both incoming and outgoing twisted vortex beams, such as such Compton scattering described by Eq.(\ref{Mt}),  we thus may define $b$ as the magnitude of the transverse displacement between the axis of a beam of matter or light and the center of a target for a straight line trajectory.  The two conjugate terms, $q(\Theta)$ and $b$, take on a different conceptual role in two opposite (delocalized vs. localized) classical limits of what we call 'wave-like' and 'particle-like' beams.  In the particle-like classical limit, corresponding to a projectile wavelength small compared to the size of an atomic target, a small projectile may have a well defined impact parameter, $\vec{b}$ with respect to the center of a diffuse target.  In the opposite wave-like limit $\vec{b}$ may be regarded as the distance of the localized atom from the axis of a broadly delocalized beam of coherent waves whose wave fronts are perpendicular to well defined rays.  In the classical particle limit $q$ may be well defined on a macroscopic scale by the scattering angle, $\Theta$. 
It is well known that in between the classical-wave and classical particle limits,  the sharp conceptual boundaries of the two classical pictures become blurred, and quantum uncertainty  appears.  Within the limits of quantum uncertainty (and/or   the classical band width theorem),  $b$ and $\Theta$ cannot both be observed with exact precision.

\section{Experimental considerations}

In this paper we concentrate on interactions of gently focused optical beams with relatively small atoms, so that $a_T << \lambda <<z_R$.  We regard the interaction as weak and fast so that first order perturbation is applicable.  Also, we restrict our attention to experiments done under single collision conditions, so that any effects of multiple beam scattering are small.  Additionally we assume that OAM and SAM transfer are distinguished by observation, particularly when $\Delta \ell = 1$.
The use of our paraxial Laguerre-Gaussian description is not valid for large vortex angles, i.e. our description is restricted to $\Theta_V(b) < 30^o$ or so\cite{paraxial}.  
Within the paraxial approximation it is sensible to regard the beam to be approximately parallel to the beam axis.
Our Gauss-Laguerre approximation, applicable to differential cross sections, is based on the paraxial approximation, i.e $\cos\Theta_V(b) \simeq 1$ and $\sin\Theta_V(b) \simeq \Theta_V(b)$.  Consequently, the straight line beam rays are approximately parallel to the beam axis, even at large $b$, and the parameter, $\vec{b}$, corresponds to that of an incoming ray along a straight line trajectory, $\vec{R} = \vec{b} + \vec{c} t $, approximately parallel to the (conventionally incoming) beam axis.   Thus to test calculations using Eq(\ref{Mt}), one may use an extended target, such as a simple gas cell of uniform density, so long as the beam center is focused near the center of the target, and the thickness of the target region is not large compared to $z_R$. 
Within this region of small $z$, the beam intensity at any fixed $b$ is nearly constant.
More general descriptions, using a Bessel \cite{MatulaEtal, IvanovSerbo, Ivanov}, a higher order Bessel\cite{Eberly}, or a higher order Mathieu\cite{mathieu} basis to describe vortex beams are available, but for us they are less convenient mathematically than the approximate Laguerre-Gaussian description we use.  But use of the Bessel basis could require different restrictions for comparison of experimental observations with calculations.   We note that similar issues have been addressed in the field of ion-atom collisions\cite{Mc,MC1}. 

We believe that Eq(\ref{Mt}) above may be applied, in some cases, to electron beams or hard x-ray beams with $\lambda < a_T$.  However, the restrictions on experimental design may differ from those above.  

A useful way to directly test the validity and accuracy of vortex atomic form factors is to design experiments to observe the difference between the absolute square of the vortex atomic form factor and its plane-wave limit, corresponding to,

\begin{eqnarray}
     T_v = \frac{ |M_v|^2 - |M_p|^2 }{|M_p|^2} =  \frac{|M_v|^2 }{|M_p|^2} -1 
   \label{T} 
\end{eqnarray} 
Here $M_p$ denotes the plane-wave wave limit of $M_v$, which corresponds to a relatively large value of $z_R$ (at fixed $\lambda$) as compared to that used for $M_v$, and $M_p$ has the same quantum numbers as $M_v$.   $T_v$ isolates the effect of the finite beam width in a vortex beam.  This vortex factor, $T_v$, may be directly measured.  This factor distinguishes vortex size effects from the plane wave limit.  Furthermore, $T_v$, may be used to convert either experimental or theoretical data for plane wave (or cylindrical) photons to data for vortex photons (or electrons) by multiplying the data by $T_v + 1$.  The factor $T_v+1$ may be used to convert the absolute square of an atomic form factor to the absolute square of a vortex atomic form factor.  Thus $T_v + 1$ is a vortex conversion factor. 

To observe $T_v$ experimentally, for any process in which $|M_v|^2$ appears as a simple factor as in Eq(\ref{dsig}),  one could measure the number of counts in a time interval, then increase $z_R$ significantly and repeat this measurement.  This has the advantage that some experimental systematic errors can cancel in the ratio. In general $T_v$ is a function of $q(\Theta)$, although for optical photons where $q a_T << 1$ the variation in $\Theta$ is expected to be small.   Also for inelastic scattering there is a minimum value of $q$ depending on the process involved, so that cross sections might yield different values of $T_v$ for different physical processes\cite{MC1,Mc5}.   It is noted again here that $M_p$ corresponds to the plane wave limit of $M_v$. 

In the field of ion-atom collisions a continuous transition from a coherent wave-like trajectory with $q a_T << 1$ to an incoherent particle-like trajectory with $q a_T >>1$ has been observed\cite{MMM}.
In either the classical wave limit or the classical particle limit, the beam trajectory may be described by a well defined trajectory, $\vec{R}(t) = \vec{b} + \vec{v} t $, where $b = d_0 b'$ with $d_0$ being a physically measurable distance scale, e.g., either $w(0)$ or $a_T$ in the context of this paper.  Here $b'$ is a dimensionless, continuously varying number that defines the value of the impact parameter, $b$, on the scale of $d_0$.   
In a full quantum description using wave-packets this dimensionless number $b'$ may be used to join the classical wave ray trajectory with the classical particle trajectory.  This is illustrated in the method of virtual impact parameters\cite{Mc3}, that may be used when the magnitude of the probability amplitude, $|a(\vec{b}')|$, varies significantly from both 0 and 1 in spatial regions of $\vec{b}'$ where the probability is diffuse, i.e., where the size of the probability cloud of the projectile is comparable to the size of the atomic target.  From this diffuse region of space without sharp classical boundaries, scattering from a range of different values of $\vec{b}'$ may scatter coherently into the same scattering angle, $\Theta$, (or $q$).  In this case one may integrate over virtual impact parameters in the region, $0 < |a(\vec{b}')| < 1$, where $a(\vec{b}')$ contributes over a range of virtual impact parameters, $\vec{b}'$, to  obtain a physically observable $|f(\vec{q})|^2$ at an experimentally well defined value of $\Theta$ . Thus, an observable quantity, which changes continuously from coherent scattering at small $q d_0$ to a incoherent scattering at large $q d_0$ may be evaluated by using a single, unifying, quantum mechanical description.

In ion-atom collisions one such observable quantity is the square of the effective charge of the projectile, $Z^2_{eff}(q)$ (expressed in units of $e^2$), in a regime where first order perturbation may be applied, over a range of $q$ between the classical wave ($q a_T << 1$) and the classical particle limits $(q a_T >>1)$.    In the coherent, small $q$ limit (where classical wave methods apply), the observable effective charge squared, $Z^2_{eff}$ of a projectile atom or ion of nuclear charge $Z$ is fully screened by its $N$ surrounding electrons so $Z^2_{eff} \rightarrow (Z-N)^2$, while in the incoherent, large $q$ limit, all the projectile charges scatter independently so that $Z_{eff} \rightarrow Z^2 + N$.  For a neutral atomic projectile $N = Z$, while $N < Z$ for a partially stripped ion.  It is the continuous change of $Z^2_{eff}(q)$ between $(Z - N)^2$ at small $q$ and $Z^2 + N$ at large $q$  that has been observed\cite{MMM, Mc3}.   We suggest that for vortex photons such an analysis might be applied to the vortex factor of Eq(\ref{T}) when $\lambda/z_R << 1$, since this condition is analogous to $Z v_0/v  <<  1$ that defines the region of validity for use of perturbation theory in ion-atom collisions, where $Z v_0$ is the speed of the electron about the nucleus of the projectile ion, and $v$ is the speed of the projectile itself.

Our results may be applied to multi-centered molecules and clusters\cite{MetcalfEtal}. Using the idea that a shift, $\vec{\cal R}$, in position space corresponds to a factor of $e^{i \vec{q} \cdot \vec{\cal R}}$ in momentum space, it is straightforward to extend use of the atomic form factor to multi-centered systems using a geometric structure factor, $G_{N_I}$. This factor has been described in detail elsewhere\cite{Mc4} within the independent center, independent electron approximation, and has been widely used for some time in molecular, condensed matter, nuclear, and high energy physics\cite{Mc4}. 

Finally we turn to some considerations in developing a computer code to evaluate the vortex atomic form factor, $M_v$, of Eq(\ref{Mt}), as well as the vortex factor of Eq(\ref{T}), for comparison to experiment.  One challenge arises in working with different symmetries (spherical for small targets and cylindrical for photon beams) that includes different directions for incoming and outgoing beams.  After considering various approaches, we recommend use of a Cartesian grid for numerical evaluation, expressing all coordinates in Eq(\ref{Mt}) above in terms of an $(x,y,z)$ coordinate system corresponding to the incoming half of the interaction.  The $(x',y',z')$ coordinates for the outgoing part (evident in Eq(\ref{Mt})) are related to the $(x,y,z)$ coordinates by a two dimensional rotation about the common $x$ axis defined by the $y-z$ plane of scattering.  The angle of rotation is the scattering angle, $\Theta $, between the incoming and outgoing vortex beams.  We also note that the parameter range, corresponding to existing conditions for twisted optical photon beams, is $z_R >> \lambda >> a_T$.   Under these conditions we expect that $M_v$ will be largely independent of $q$, since $q a_T << 1$.  When $\lambda << z_R$, we expect that $T_v$ should vary as $\frac{1}{\lambda z_R}$.   In the future a code might be available upon request\cite{PG}.    

\section{Summary}

In summary we have considered both mathematical evaluation and experimental observation related to the atomic form factor for twisted (vortex) photon beams interacting with atomic targets.  We have resolved three problems specified in an earlier paper that gave a detailed derivation of a formula for this vortex atomic form factor.  First we properly normalized the expression for the scale-invariant form factor and explained how area and volume normalizations differ, but lead to the same physical observables.  Second we examined the effect of moving the target atom away from the center of the incoming vortex beam.  Summing count rates in differential cross sections over impact parameters is related to summing over corresponding cross sections differential in momentum transfers (or scattering angles).  Both yield identical total cross sections.  
While relatively small particles can interact with well separated atoms one at a time, relatively large waves can interact with many target atoms.  Understanding this enabled us to suggest experiments, for both total and differential cross sections, using macroscopic gas cells.  This understanding also eliminates any special need, other than including the $b$-dependence arising from the $q$-dependence in the atomic form factor, to account for the effect of moving target atoms away from the exact center of a macroscopic vortex beam.  Third we formulated photon-atom scattering in an impact parameter (position) representation and related this to a complementary momentum transfer representation.  
The momentum representation precisely describes a large coherent classical beam of photons interacting with many atoms in a macroscopic target, while the position representation is well suited to describe the classical particle limit of a single, small photon interacting with a single atom.   The method of virtual impact parameters might be useful in the quantum regime between these two classical limits.

Possible new experiments observing ratios of counting rates for vortex photons interacting with atomic matter, and their plane wave limits were considered.  Such experiments could usefully test the accuracy of tables of conversion factors that relate a variety of photon-matter interaction processes (e.g. that involve transfer of photonic orbital angular momentum) using  vortex beams.
A convenient, measurable, vortex factor can be simply expressed in terms of the absolute square of the vortex atomic form factor and its asymptotic plane wave limit.

\section{Acknowledgements}

JHM  gratefully acknowledges useful discussions with A. Salin, C. L. Cocke, W. Thompson, J. Eberly, B. Davis, and M. Frow.   This work was supported in part by the NSF under Grants No. PHY-1005709 and PHY-1205788.

\newpage

{\bf Figure Captions:}

\vskip 2em

Figure 1.  Envelope of classical light rays associated with a twisted vortex beam at a radial distance  $\rho(z) = w(z)$ from the beam axis.  At $ z=0$ all segments shown have the same distance of closest approach, $b$, to the beam center, i.e., $b \equiv \rho(0)  = w(0) = \sqrt{\lambda z_R/\pi} $.  Each segment shown is a straight line.  The scale for $w(z)$ is the same as in Fig 2.

\vskip 2em

Figure 2.  Side view of envelopes of intensity for a Gaussian beam. Each envelope shown corresponds to a surface of constant beam intensity, ranging from maximum intensity along the center line to the intensity that has decreased exponentially by a factor of $\frac{1}{e^2} \simeq 0.135$.  A three dimensional view of the outermost envelope shown here is given in Fig. 1.  
Envelopes of different intensity may not cross as then the intensities would not then be different. 
Each envelope shown here corresponds to a ring of classical light rays of the same intensity, as shown in Fig. 1, but at a distance $\rho(z) = b \frac{w(z)}{w(0)}$ from the centerline, i.e. a distance reduced by a factor of $b/w(0)$ from that shown in Fig 1. Thus, as $b$ decreases, the intensity of the rays within this envelope increases.
The horizontal scale of $z$ is in units of the Rayleigh range, $z_R$.  
The vertical scale of  $\rho(z) = b \sqrt{1 + z^2}$ is given in units of $w(0)$ for $0 \leq b \leq w(0)$.
Fig 2B shows the paraxial region near the center of the beam where envelopes of the same beam intensity are nearly parallel to the beam axis.  Fig 2C shows the asymptotic region, far from the beam center, where envelopes of the same beam intensity are well described by cones corresponding to Eq(\ref{b-Thetav}).  
The single envelope shown in Fig 1 corresponds to $b = w(0) = \sqrt{\lambda z_R/\pi}$.


\vskip 2em


\section{References}

\begin{thebibliography}{99}


\bibitem{Allen92}  L Allen, M W Beijersbergen, R J C Spreeuw, and J P Woerdman, Phys. Rev. Lett.{\bf 45}, 8185 (1992).

\bibitem{YaoPadgett} Alison M. Yao and Miles J. Padgett, Advances in Optics and Photonics \textbf{3}, 161-204 (2011).

\bibitem{TU} http://lamp.tu-graz.ac.at/~hadley/ss1/crystaldiffraction/atomicformfactors/formfactors.php  

\bibitem{Chantler00} C T Chantler,  J. Phys. Chem. Ref. Data 29, 597 (2000).  

\bibitem{nist}   www.nist.gov/data/PDFfiles/jpcrd67.pdf‎   

\bibitem{GarveyGreen76} R H Garvey and A E S Green,  Phys. Rev. {\bf A 13}, 931 (1976).  

\bibitem{AfanasevEtal}  Andrei Afanasev, Carl E. Carlson, and Asmita Mukherjee, Phys Rev {\bf A 88} 033841 (2013).  

\bibitem{MatulaEtal} O Matula, A G Hayrapetyan, V G Serbo, A Surzhykov and S Fritzsche,  J. Phys. B {\bf 46} (2013) 205002.  

\bibitem{DKM}  Basil S Davis, L Kaplan and J H McGuire, J. Opt. {\bf 15} (2013) 035403; corrigendum:  J. Opt. 15 (2013) 109501.  

\bibitem{Beth} R. A. Beth, Phys. Rev. \textbf{50}, 115 (1936).

\bibitem{Molina-TerrizaTorresTorner} Gabriel Molina-Terriza, Juan P. Torres, and Lluis Torner, Nat. Phys. {\bf 3} 305 (2007).  

\bibitem{BabikerBADR} M. Babiker, C. R. Bennett, D. L. Andrews, and L. C. D\'{a}vila Romero, Phys. Rev. Lett. {\bf 89}, 143601 (2002). 

\bibitem{MolinaTerrizaRTTW} G. Molina-Terriza, J. Recolons, J. P. Torres, L. Torner, and E. M. Wright, Phys. Rev. Lett. {\bf 87}, 023902 (2001).

\bibitem{BerryMcDonald} M. V. Berry and K. T. McDonald, J. Opt. A \textbf{10}, 035005 (2008).

\bibitem{GiovanniniNMS} Daniele Giovannini, Eleonora Nagali, Lorenzo Marrucci, and Fabio Sciarrino, Phys. Rev. A {\bf 83} 042338 (2011). 

\bibitem{IvanovSerbo} I. P. Ivanov and V. G. Serbo,  Phys Rev {\bf A 84} 033804 (2011).   

\bibitem{Ivanov} I. P. Ivanov, Phys Rev {\bf D 83} 093001 (2013).  

\bibitem{TorresTornerBook}   {\it Twisted Photons: Applications of Light with Orbital Angular Momentum}, (Wiley- VCH, 2011, ISBN: 978-3-527-40907-5),  Juan P Torres and Lluis Torner, editors.

\bibitem{WildeUskov} M. M. Wilde and D. B. Uskov, Phys. Rev. A \textbf{79}, 022305 (2009).  

\bibitem{BarreiroWeiKwiat} J. T. Barreiro, T.-C. Wei, and P. G. Kwiat, Nat. Phys. {\bf 4} 282 (2008).   

\bibitem{MairVWZ} Alois Mair, Alipasha Vaziri, Gregor Weihs, and Anton Zeilinger, Nature {\bf 412} 313 (2001).

\bibitem{Bucks}  P. H. Bucksbaum, Physics Today  {\bf 59}, 57 (2006).  

\bibitem{Shakovetal}  Kh. Kh. Shakov and J. H. McGuire, Phys. Rev. A {\bf 67}, 033405 (2003).   

\bibitem{PadgettBowman}  Miles Padgett and Richard Bowman, Nature Photon {\bf 5} 343 (2011)  

\bibitem{MTetal}  G Molina-Terriza, A Vaziri, R Ursin, and A Zeilinger, Phys Rev Lett {\bf 94} 040501 (2005).  

\bibitem{AltmanEtal}  A R Altman, K G K\"opr\"ul\"u, E Corndorf, P Kumar, and G A Barbosa, Phys. Rev. Lett. {\bf 94} 123601 (2005).  

\bibitem{AndersenEtal} M. F. Andersen, C. Ryu, Pierre Clad\'e, Vasant Natarajan, A. Virizi, K. Helmerson and W. D. Phillips, Phys Rev Lett {\bf 97} 170406 (2006).    

\bibitem{WikiAF}  http://en.wikipedia.org/wiki/Atomic\_form\_factor   

\bibitem{Mc}  J. H. McGuire, \textit{Electron Correlation Dynamics in Atomic Collisions} (Cambridge, UK: Cambridge University Press, 1997), Chapter 9.  

\bibitem{Heitler} W. Heitler, \textit{The Quantum Theory of Radiation,} Third Edition (Oxford: Clarendon Press, 1954).  

\bibitem{Mc1} See Ref.~\cite{Mc}, Eq. (9.41). 

\bibitem{KleinNishina}  http://en.wikipedia.org/wiki/Klein–Nishina\_formula  

\bibitem{Econs}  This is standard in scattering theory.  See D E Soper,     http://physics.uoregon.edu/~soper/QuantumMechanics2006/Smatrix.pdf   
Note the discussion around Eq(17) (noting that we set $e^{ i (E_F - E_I) t} \simeq 1$ in our present development). 

\bibitem{MC1} M. R. C. McDowell and J. P. Coleman, {\it Introduction to the Theory of Ion-Atom Collisions}, (North Holland, N.Y., 1970)     

\bibitem{ArfkinWeber} George B  Arfkin and Hans J  Weber, {\it Mathematical Methods for Physicists} (Academic Press)  Chapter on Fourier Transforms.    

\bibitem{Mc1a}  See Ref.~\cite{MC1}, Chapter 1.   

\bibitem{Golds}  Herbert Goldstein, {\it Classical Mechanics} (Addison Wesley).   

\bibitem{Mc2}   See Ref.~\cite{Mc}, Appendix A.4.   

\bibitem{MS}  R. McCarroll and A. Salin, Compt. Rend. Acad. Sci. {\bf 263}, 329 (1966).  

\bibitem{vortexClass} We have used the asymptotic relation for Gaussian beams (Ref.~\cite{paraxial}) for $z >> z_R$.   In our paper $\Theta_{\cal V}$ is the same as the divergence of the beam, $\theta$ in Ref.~\cite{paraxial} (but the meanings of $b$ differ).   


\bibitem{paraxial}  http://en.wikipedia.org/wiki/Gaussian\_beam  \   

\bibitem{Eberly}  J. Durnin, J. J. Miceli Jr., and J. Eberly, Phys Rev Lett {\bf 58} 1499 (1987). 

\bibitem{mathieu} J. C. Gutierrez-Vega, M. D. Iturbe-Castillo, and S. Ch\'avez-Cerda,  Opt. Lett. {\bf 25} 1493 (2000).  

\bibitem{Mc5}  In ref.~\cite{Mc}, see the equation below Eq(2.32) and also the nearby equations in Sec. 2.2 that depend on $q_\parallel = \Delta E/v$, which is non-zero for inelastic collisions.

\bibitem{MMM}  E C Montenegro, W E Meyerhof and J H McGuire, {\it Adv. in At. Mo. and Optical Phys.} {\bf 34} 250 (1994).

\bibitem{Mc3}  Section 8.2.2 of reference\cite{Mc}; J Wang, J H McGuire and E C Montenegro, Phys. Rev. {\bf A 51} 504 (1995). 

\bibitem{MetcalfEtal}  Stefan Evans, John Noé, Harold Metcalf,  http://laser.physics.sunysb.edu/~stefan/journal/journal.html  (2013). 

\bibitem{Mc4}  See ref.~\cite{Mc}, Section 4.2.  

\bibitem{PG}  Pierson Guthrey, private communication, presently in Applied Mathematics, Iowa State University, Ames, IA 50011.   

\end{thebibliography}
\end{document}